\documentclass[11pt,A4]{article}
\usepackage[latin1]{inputenc}
\usepackage{amssymb}
\usepackage{amsmath}
\usepackage{amsfonts}
\usepackage{amsbsy}
\usepackage{natbib}
\usepackage{eucal}
\usepackage{graphicx}
\usepackage{titlesec}
\usepackage[english]{babel}
\usepackage{times}
\usepackage{titling}
\usepackage{mathptmx}
\usepackage{float}
\usepackage[font=small]{caption}
\usepackage{array}
\usepackage[colorlinks,citecolor=blue,urlcolor=blue]{hyperref}
\usepackage{authblk}
\usepackage{amsthm}
\usepackage{subfig} 
\usepackage{xcolor}

\usepackage{setspace}
\DeclareSymbolFont{txgreek}{OML}{cmr}{m}{it}

\renewcommand{\abstract}[1]{{\small\noindent
\hrulefill\par \vspace*{0.1cm}\noindent{\small\bf\sffamily
{Abstract}}\parindent=0pt\par\noindent\vspace{-0.1cm}\noindent\hrulefill\par\vspace*{0.5\baselineskip}\hspace*{0cm}\renewcommand{\baselinestretch}{1.1}\sffamily{#1}\par\vspace*{-0.1cm}\noindent\hrulefill}}

\def\and{,\;}
\DeclareMathSizes{11}{11}{7.8}{6.5}

\def\paragraf{\fontsize{9}{10pt}\fontfamily{phv}\fontshape{it}\selectfont}
\titleformat{\paragraph}
{\titlerule[0pt]
\vspace{0cm}%
\paragraf}
{\theparagraph}{.5em}{\vspace*{0\baselineskip}}

\def\titol{\fontsize{12.045}{12pt}\fontfamily{phv}\fontseries{b}\selectfont}
\titleformat{\section}
{\titlerule[0pt]
\vspace{0.2cm}%
\titol}
{\thesection.}{.5em}{\vspace*{0\baselineskip}}

\def\titolp{\fontsize{11.045}{11pt}\fontfamily{phv}\fontseries{b}\fontshape{it}\selectfont}
\titleformat{\subsection}
{\titlerule[0pt]\bigskip
\vspace{-0.4cm}%
\titolp}
{\thesubsection.}{.5em}{\vspace*{0\baselineskip}}

\def\titolpp{\fontsize{10.045}{10pt}\fontfamily{phv}\fontshape{it}\selectfont}
\titleformat{\subsubsection}
{\titlerule[0pt]
\vspace{-0.4cm}%
\titolpp}
{\thesubsubsection.}{.5em}{\vspace*{0\baselineskip}}

\usepackage{titling}
    \pretitle{\begin{center}\sffamily\fontsize{18pt}{20pt}\selectfont}
    \posttitle{\par\end{center}}
    \preauthor{\begin{center}\fontsize{12pt}{14pt}\selectfont}
    \postauthor{\par\end{center}\vspace{0bp}}
    \predate{}
    \date{}
    \postdate{}

\title{Finding archetypal patterns for binary questionnaires}

\thanksmarkseries{arabic}
\author{Ismael Cabero\thanks{Depart\textcolor{black}{a}ment de Did\`actica de les  Matem\`atiques, Universitat de Val\`encia, Spain } \and  Irene Epifanio\thanks{Departament de Matem\`atiques-IMAC\textcolor{black}{-IF}, Universitat Jaume I, Castell\'o 12071, Spain. Email: epifanio@uji.es} }

\def\headers#1{\fontsize{8.5}{10}\centering\sffamily\itshape{#1}}
\def\page#1{\fontsize{8.5}{10}\sffamily{#1}}
\usepackage{fancyhdr}

\begin{document}
\maketitle

\thispagestyle{empty}
\renewcommand{\headrulewidth}{0truecm}
\pagestyle{fancy}
\rhead[\headers{Finding archetypal patterns for binary questionnaires}]{\page{\thepage}}
\lhead[\page{\thepage}]{\headers{I. Cabero and I. Epifanio}}
 \lfoot{} \rfoot{}
\cfoot{}

\abstract{Archetypal analysis is an exploratory tool that explains a set of observations as mixtures
of pure (extreme) patterns.If the patterns are actual observations of the sample,
we refer to them as archetypoids. \textcolor{black}{For the first time, w}e propose to use archetypoid analysis for binary
observations. This tool can contribute to the understanding of a binary data set, as in
the multivariate case. We illustrate the advantages of the
proposed methodology in a simulation study and two applications, one exploring objects (rows) and the other exploring items (columns). One is related to determining student skill set profiles and the other to describing item response functions.}

\paragraph{MSC:} \vspace{-0.5cm} 62H99, 62P25\textcolor{black}{, 97D60}

\paragraph{Keywords:} \vspace{-0.5cm} dichotomous  item test, archetypal analysis, functional data analysis, item response theory, skill profile

\renewcommand{\baselinestretch}{1.2}
\bigskip

\section{Introduction} \label{introduccion} Mining binary survey data is of utmost importance in social sciences. 
Many raw data from exams, opinion surveys, attitude questionnaires, etc. come in the form of a binary data matrix, i.e. examinees' responses are coded as 0/1 (1 if examinee $i$ answers item $h$ correctly, otherwise 0). \textcolor{black}{The binary matrix can be viewed from two points of views. In the first, the interest lies in the rows, i.e. in the people, while in the second, the interest lies in the columns that contain the items or variables. In both cases, exploratory data analysis (EDA) aims to find information in data and generate ideas \citep{UNWIN2010156}. In order to be useful as a tool for EDA on data sets, the tool should be simple and easy to use, with few parameters, and reveal the salient features of the data in such a way that humans can visualize them \citep{1672644}.}

For the first time, we propose the use of \textcolor{black}{the exploratory tool} Archetypoid Analysis (ADA) for  this kind of data in order to understand, describe, visualize and extract information that is easily interpretable, even by non-experts. ADA is an unsupervised statistical learning technique (see \citet[Chapter 14]{HTF09} for a complete review of unsupervised learning techniques). Its objective is to approximate sample data by a convex combination (a mixture) of $k$ pure patterns, the archetypoids, which are extreme representative observations of the sample. Being part of the sample makes them interpretable, but also  being extreme cases facilitates comprehension of the data. Humans understand the data better when the observations are shown through their extreme constituents \citep{Davis2010} or when features of one observation are shown as opposed to those of another \citep{Thurau12}.

ADA was proposed by \cite{Vinue15} \textcolor{black}{for real continuous multivariate data} as a derivative methodology of Archetype Analysis (AA). AA was formulated by \cite{Cutler1994}, and like ADA, it seeks to approximate data through mixtures of archetypes\textcolor{black}{, also for real continous multivariate data}. However, archetypes are not actual cases, but rather a mixture of data points. Recently, \cite{Eugster14} proposed a probabilistic framework of AA (PAA) to
accommodate binary observations by working in the parameter space.

 AA and ADA have been applied to many different fields, such as astrophysics \citep{Chan2003}, biology \citep{Esposito2012}, climate \citep{doi:10.1175/JCLI-D-15-0340.1,w9110873}, developmental
psychology \citep{SAM16}, e-learning \citep{Theodosiou}, finance \citep{Moliner2018a}, genetics \citep{Morup2013}, human development \citep{Epifanio2016,doi:10.1080/00031305.2018.1545700}, industrial engineering \citep{EpiVinAle,EpiIbSi17,MillanEpi,10.1371/journal.pone.0228016}, machine learning \citep{Morup2012,EugsterPAMI,Eugster14,Rago15,CabEpi19}, market research \citep{Li2003,Porzio2008,Midgley2013}, multi-document summarization \citep{Canhasi13,Canhasi14}, nanotechnology \citep{doi:10.1021/acsnano.5b05788}, neuroscience \citep{griegos,Morup16} and sports \citep{Eugster2012,VinEpi17,VinEpi19}.

Archetypal analysis techniques lie somewhere in between two well-known unsupervised statistical techniques: Principal Component Analysis (PCA) and Cluster Analysis (CLA). In data decomposition techniques, a data set is viewed as a linear combination of several factors to find the latent components. Different prototypical analysis tools arise depending on the constraints on the factors and how they are combined \citep{Morup2012,Vinue15}.  The factors with the least restrictions are those produced by PCA, since they are  linear combinations of variables. One of the advantages is that this helps explain the variability of the data; however, the interpretability of the factors is compromised. Instead, the greatest restrictions are found in cluster  tools, such as $k$-means or $k$-medoids. Their factors are readily interpreted because they are centroids (means of groups of data) or medoids (concrete observations) in the case of $k$-means and $k$-medoids, respectively. The price that clustering tools pay for interpretability is their modeling flexibility due to  the binary assignment of data to the clusters. Archetypal tools, on the other hand, enjoy higher modeling flexibility than cluster tools but without losing the interpretability of their factors. A  table summarizing the relationship between several unsupervised multivariate  techniques is provided by \cite{Morup2012} and \cite{Vinue15}.

AA and ADA  were originally thought of for real-valued observations. \textcolor{black}{The aim of this work is to extend archetypal tools to binary data.} For AA,
as the factors (archetypes) are a mixture of data, they would not necessarily be binary vectors, and as a consequence they would not be interpretable. In ADA though, the factors (archetypoids) are actual cases, so ADA can be applied to binary data without losing the  interpretability of the factors. \textcolor{black}{So, among the possible archetypal techniques (AA, PAA and ADA), we propose to use ADA for binary data.}

To perform a sanity check and provide insight we analyze the solutions obtained by AA, PAA and ADA through a simulation study\textcolor{black}{, where ADA shows its appropriateness versus AA or PAA for binary data sets}. Furthermore, we present two real applications and compare ADA solutions with those of other established unsupervised techniques to illustrate the advantages of ADA in educational and behavioral
sciences, when used as another useful tool for data mining in these fields \citep{reviewedu}. \textcolor{black}{In the first application, we are interested in rows, while in the second application in columns.}

The outline of the paper is as follows: In Section \ref{metodologia} we review AA and ADA for real-valued multivariate and functional data and PAA\textcolor{black}{, besides other multivariate techniques used in the comparison}. In Section \ref{binario} we 
introduce the analysis for binary multivariate data.  In Section \ref{simulacion}, 
a simulation study with binary data compares the different strategies for obtaining archetypal patterns.
  In Section \ref{aplicaciones}, our proposal is applied to two real data sets and compared to the results of other well-known unsupervised statistical learning techniques. 
Section \ref{conclusiones} contains conclusions and some ideas for future work. 

The data sets and code in R \citep{R} for reproducing the results for both artificial and real data are available at \url{http://www3.uji.es/~epifanio/RESEARCH/adaedu.rar}.

\vspace{-0.8cm}

\textcolor{black}{\section{Preliminary} \label{metodologia}}
\subsection{AA and ADA in the real-valued multivariate case} 
Let $\mathbf{X}$ be an $n \times m$ real-valued matrix with $n$ observations and $m$ variables. Three matrices are established in AA: a) the $k$ archetypes $\mathbf{z}_j$, which are the rows of a $k \times m$ matrix $\mathbf{Z}$; b) an $n \times k$ matrix $\mathbf{\alpha} = (\alpha_{ij})$ with the mixture coefficients that approximate each observation $\mathbf{x}_i$ by a mixture of the archetypes  ($\mathbf{\hat{x}}_i = \displaystyle \sum_{j=1}^k \alpha_{ij} \mathbf{z}_j$); and c) a $k \times n$ matrix $\mathbf{\beta} = (\beta_{jl})$ with the mixture coefficients that characterize each archetype ($\mathbf{z}_j$ = $\sum_{l=1}^n \beta_{jl} \mathbf{x}_l$). To figure out these matrices, we minimize the following residual sum of squares (RSS) with the respective constraints ($\| \cdot\|$ denotes the Euclidean norm for vectors):

\begin{equation} \label{RSSar}
RSS = \displaystyle  \sum_{i=1}^n \| \mathbf{x}_i - \sum_{j=1}^k \alpha_{ij} \mathbf{z}_j\|^2 = \sum_{i=1}^n \| \mathbf{x}_i - \sum_{j=1}^k \alpha_{ij} \sum_{l=1}^n \beta_{jl} \mathbf{x}_l\|^2{,}
\end{equation}

under the constraints

\begin{enumerate}

\item[1)] $\displaystyle \sum_{j=1}^k \alpha_{ij} = 1$ with $\alpha_{ij} \geq 0$ {for} $i=1,\ldots,n$ {and}

\item[2)] $\displaystyle \sum_{l=1}^n \beta_{jl} = 1$ with $\beta_{jl} \geq 0$ {for} $j=1,\ldots,k${.}

\end{enumerate}

As previously mentioned, archetypes do not necessarily match real observations. Indeed, this will only happen when one and only one $\beta_{jl}$ is equal to one for each archetype, i.e. when each archetype is defined by only one observation. So, in ADA the previous constraint 2) is substituted by the following one, and as a consequence in ADA a mixed-integer problem is optimized instead of the AA continuous optimization problem:

\begin{enumerate}

\item[2)] $\displaystyle \sum_{l=1}^n \beta_{jl} = 1$ with $\beta_{jl} \in \{0,1\}$ and $j=1,\ldots,k$.
\end{enumerate}

As regards the location of archetypes, they are on the boundary of the convex hull of the data if $k$ $>$ 1 (see \cite{Cutler1994}), although this does not necessarily happen for archetypoids (see \cite{Vinue15}). Nonetheless, the archetype is equal to the mean and to the medoid in case of the archetypoid (\cite{Kaufman90}),  if $k$ = 1.

\textcolor{black}{We want to emphasize that archetypal analysis is an EDA technique based on a geometric formulation (no distribution of data is assumed). It is not an inferential statistical technique, i.e. it is not about fitting models, parameter estimation, or testing hypotheses. Nevertheless, a field to study in the future would be to view archetypal analysis as a feature extraction method \citep[Ch. 5]{HTF09}, where the raw data are preprocessed and described by $\mathbf{\alpha}$, which can be used as inputs into any learning procedure for compositional data \citep{pawlowsky2015modeling}.}

\textcolor{black}{\subsubsection{Computation of AA and ADA \label{ADA al}}}

The estimation of the matrices in the AA problem can be achieved by means of an alternating minimizing algorithm developed by \cite{Cutler1994}, where  the best $\mathbf{\alpha}$ for given archetypes $\mathbf{Z}$ and the best archetypes $\mathbf{Z}$ for a given $\mathbf{\alpha}$ are computed by turns.  To solve the convex least squares problems,  a 
penalized version of
the non-negative least squares algorithm by \cite{Lawson74} is used. \cite{Eugster2009} implemented that algorithm in the R package {\bf archetypes}, although with some changes. Specifically, the data are standardized and the spectral norm in equation \ref{RSSar} is used instead of  Frobenius norm for matrices. In our R implementation those changes were annulled, i.e. the data are not standardized by default and the objective function to minimize is defined by equation \ref{RSSar}.

With respect to the estimation of the matrices in the ADA problem, it can be achieved using the algorithm developed by \cite{Vinue15}. It is composed of two steps:  the BUILD step and the SWAP step. The objective of the BUILD step is to determine an initial set of archetypoids that will be upgraded during the following step. The objective of the SWAP step is to improve the primitive set by exchanging the selected instances for unselected observations and  checking whether these replacements decrease the RSS. \cite{JSSv077i06} implemented that algorithm in the R package {\bf Anthropometry} with three possible original sets in the BUILD step:  $cand_{ns}$,  {$cand_{\alpha}$} and  {$cand_{\beta}$}. These sets correspond  to  the nearest neighbor observations in Euclidean distance to the $k$ archetypes, the cases with the maximum $\alpha$ value for each archetype $j$ and the observations with the maximum $\beta$ value for each archetype $j$, respectively. Then three possible solutions are obtained once these three sets go through the SWAP step, but only the solution  with lowest RSS (often the same final set is returned from the three initializations) is chosen as the ADA solution.

One important point is the selection of $k$, since archetypes are not necessarily nested and neither are archetypoids. If the user has prior knowledge of the structure of the data, the value of $k$ can be chosen based on that information. Otherwise, a simple but effective heuristic \citep{Cutler1994,Eugster2009,Vinue15,Eugster14} such  as the elbow criterion can be used. With the elbow criterion, we plot the RSS
 for different $k$ values and the value of $k$ is selected as the point where the elbow is located.
 
\vspace{0.2cm}

\begin{figure}[!hbt]
\centering
  \subfloat[]{\includegraphics[width=0.4\textwidth]{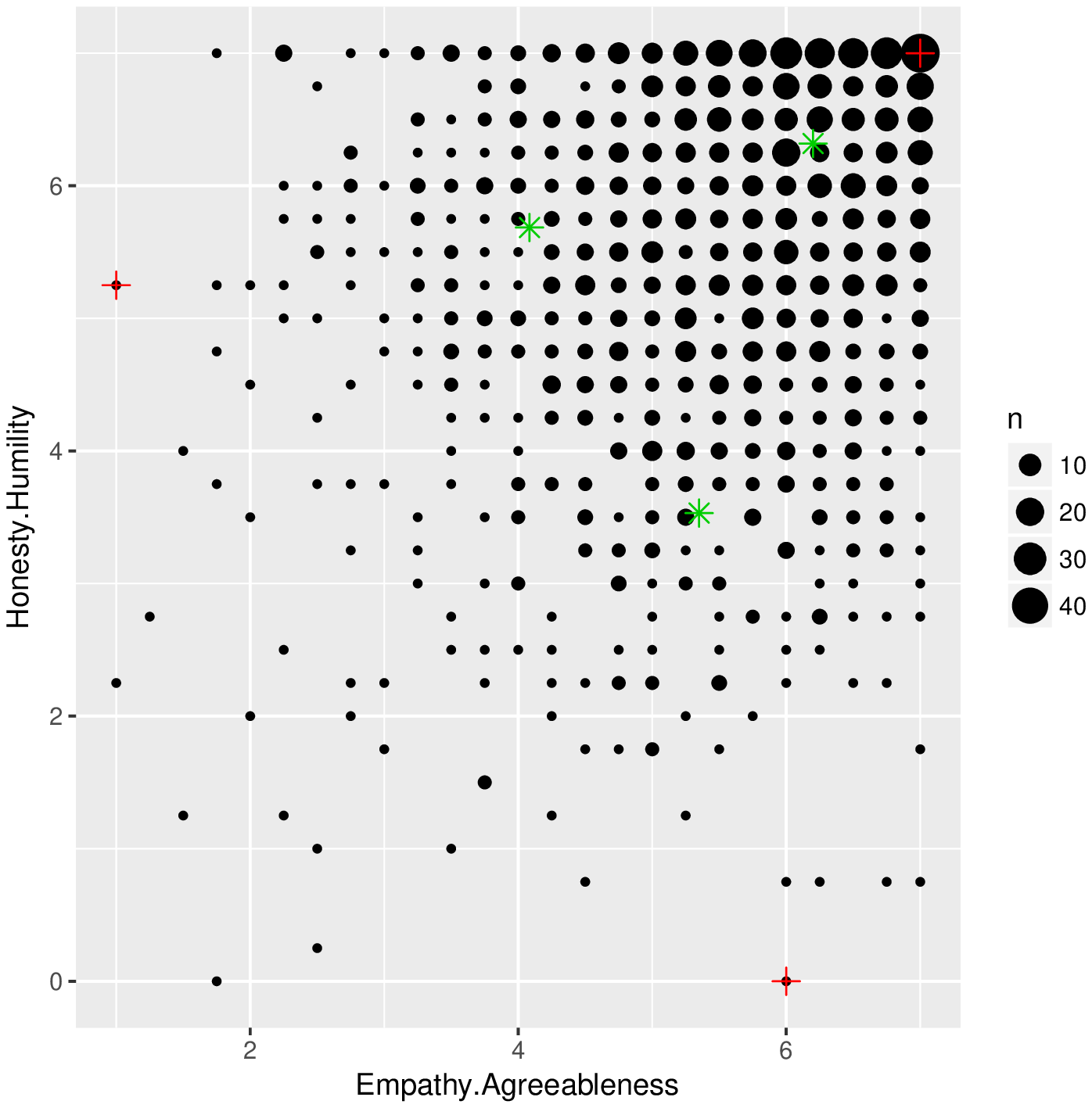}}
  \subfloat[]{\includegraphics[width=0.4\textwidth]{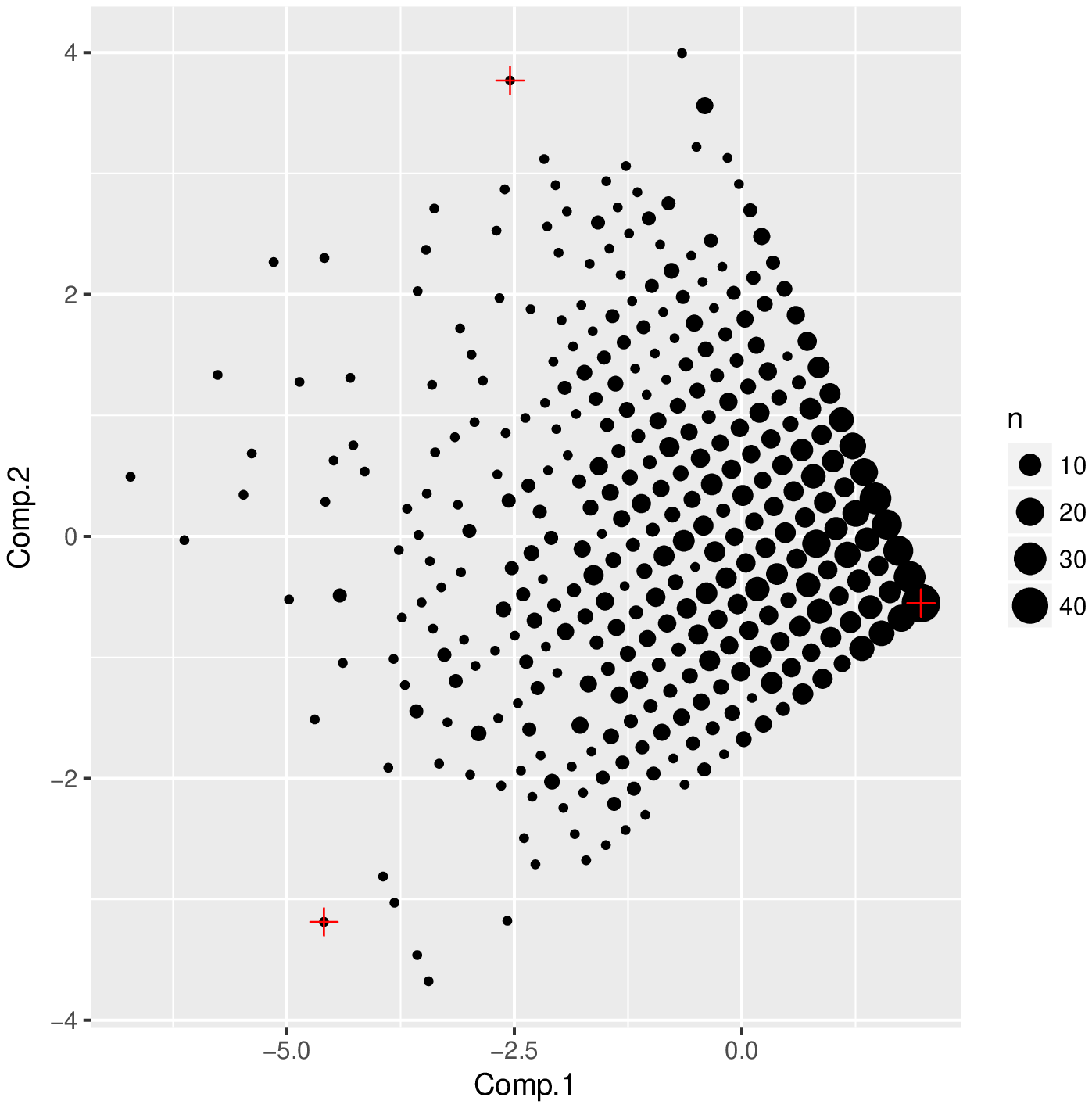}}\\
  
   \subfloat[]{\includegraphics[width=0.4\textwidth]{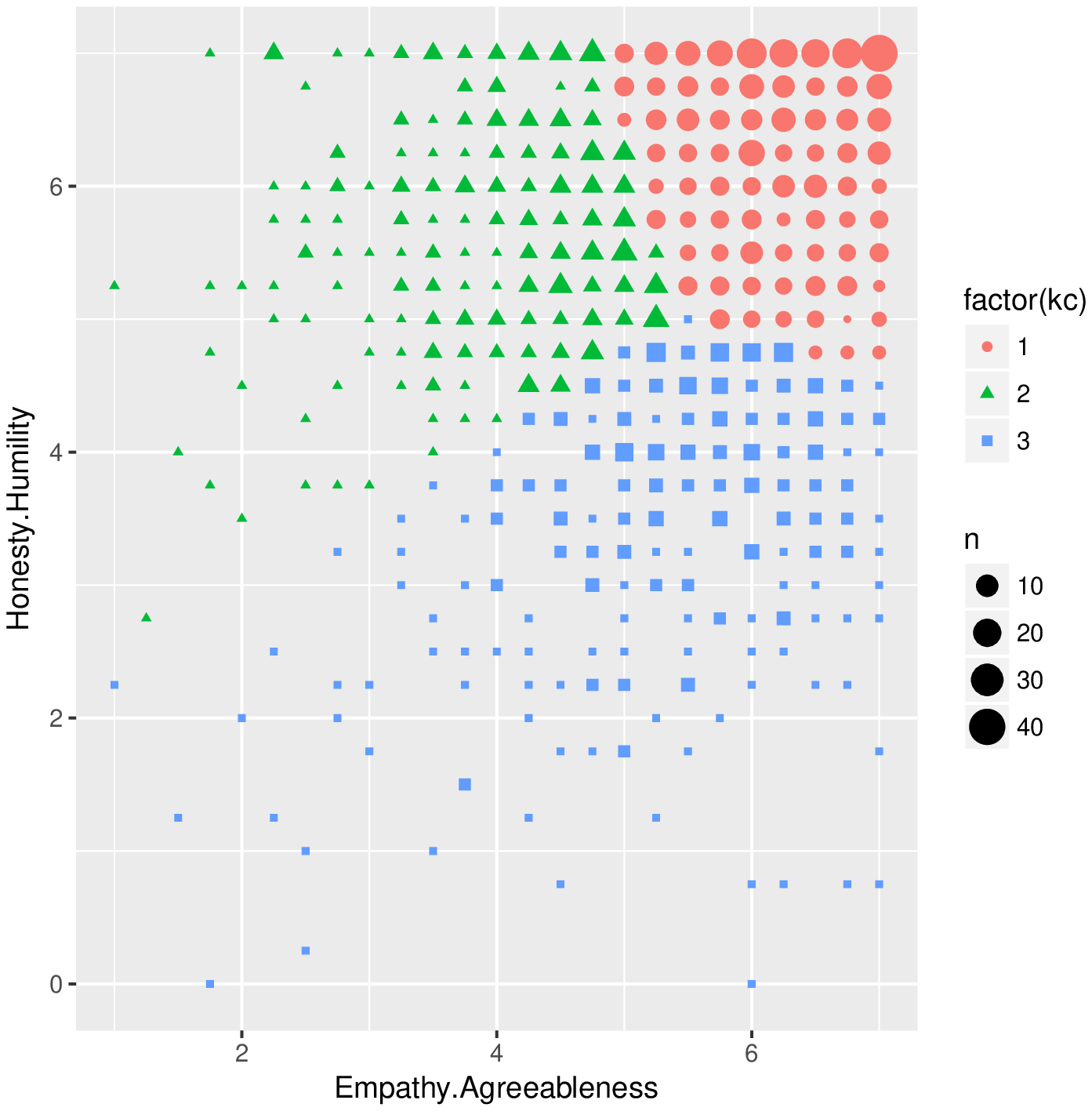}}
  \subfloat[]{\includegraphics[width=0.4\textwidth]{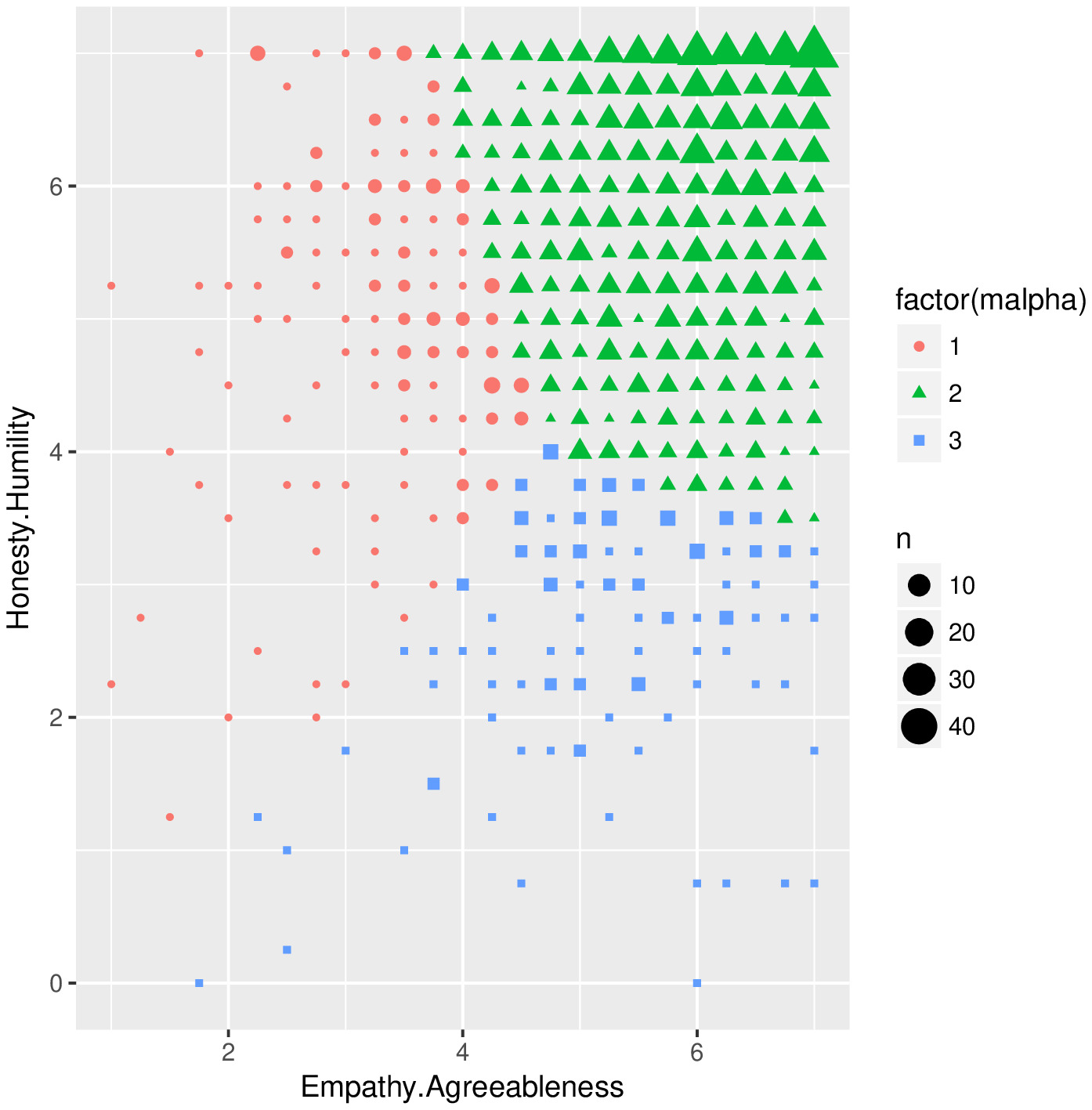}}
\caption{\label{toy} \textcolor{black}{(a) Plot of the toy example. The size of the points depends on their frequency. The red crosses represent the archetypoids, while the green stars represent the centroids of each cluster; (b) PC scores. Projected archetypoids are represented by red crosses; (c) $k$-means cluster assignments; (d) ADA assignments by the maximum alpha, i.e. assigned to the archetypoid that best explains the corresponding observation.} }
\end{figure}

\textcolor{black}{\subsubsection{Illustrative example}}
\textcolor{black}{In Figure \ref{toy} a toy two-dimensional data set is used to illustrate what archetypoids mean and the differences compared with CLA and PCA, as well as to provide
some intuition on what these pure and extreme patterns imply in behavioral sciences. Two numeric variables are considered from the data set personality-1.0 of the R package {\bf neuropsychology} \citep{neuropsychology}, which contains personality traits data from an online questionnaire: Empathy.Agreeableness and Honesty.Humility. We apply $k$-means and ADA with $k$ = 3, i.e. we find 3 clusters and archetypoids. We also apply PCA. Archetypoids are people with extreme values, which have clear profiles: archetypoid 1 is characterized by a very low Empathy.Agreeableness value together with a high Honesty.Humility value (1, 5.25), archetypoid 2 has the maximum values for both Empathy.Agreeableness and Honesty.Humility (7,7), while the third archetypoid has a very high Empathy.Agreeableness value together with the lowest  Honesty.Humility value (6,0). Archetypoids are the purest people. The rest of the individuals are expressed as mixtures (collected in alpha coefficients) of these ideal people.  For example, an individual with values of 6.25 and 0.75 for Empathy.Agreeableness and Honesty.Humility, respectively, is explained by 11\% of archetypoid 2 plus 89\% of archetypoid 3.} 

\textcolor{black}{This is compatible with the natural tendency of humans to represent a group
of objects by its extreme elements \citep{Davis2010}. Figure \ref{toy} d) shows the partition of the set generated by assigning the observations to the archetypoid that best explains each individual. However, when we apply $k$-means to this kind of data set, without differentiated clusters, the centroids are in the middle of the data cloud. Centroid profiles are not as differentiated from each other as archetypoid profiles. This happens because centroids have to cover the set in such a way that the set  is partitioned by minimizing the distance
with  respect  to the assigned centroid (see \cite{7795738} about the connection  between set  partitioning  and  clustering). On the one hand, this means that the set partition generated by $k$-means and ADA would be different (Figures \ref{toy} c) and d)). On the other hand, centroids are not the purest, and therefore their profiles are not as clear as those of archetypoids. For example, centroids 2 and 3 have values (4.1, 5.7) and (5.3, 3.5), which are not as intuitively interpretable as archetypoids. If we look again at the individual with values (6.25, 0.75) from the clustering point of view this individual is clearly assigned  to cluster 3, with centroid (5.3, 3.5), but clustering does not say anything about the distance of this point with respect to the assigned centroid, or in which direction they are separated. In fact,  (6.25, 0.75) is quite far from (5.3, 3.5). This happens because the objective of clustering is to assign the data to groups, not to explain the structure of the data  more qualitatively. Finally, note that archetypoids do not coincide with the individuals with the most extreme PC scores (see Figure \ref{toy} b)).}

\textcolor{black}{In summary, depending on our objective, the appropriate analysis should be selected. The objective of PCA  is to reduce data dimension. Although PCA returns the location of the observations in the new dimensions by PC scores, there is no 
guarantee that the principal components are interpretable. In other words, observations are expressed in a new base, but in general the PCA base is not easily interpretable. However, the objective of CLA  is to segment the data, i.e. to make groups of data by finding modes in data. Although the modes can be easily interpretable, CLA does not return an expression about the location of each observation with respect to each mode. On the other hand, finding extreme profiles, which are easily interpretable, is not the objective of PCA or CLA, but that of AA or ADA. These techniques also return the location of the observations as a function of the extreme profiles, in fact as a mixture (a convex combination), which is more easily interpretable than a linear combination. This provides a complete overview of the data set, generally supported by visual methods, i.e. this allows data to tell us more beyond the formal modeling or hypothesis testing task. }\\


\subsection{Probabilistic archetype analysis}
The idea underlying PAA is to work in a parameter space instead of the observation space, since the parameter space is often vectorial even if the sample space is not.  The key is to assume that data points come from a certain distribution (from the Bernoulli distribution in the case of binary observations). Then the maximum likelihood estimates of the parameters of the distributions are seen as the parametric profiles that best describe each observation, and archetypal profiles are computed in the parameter space by maximizing the corresponding log-likelihood under the constraints for $\mathbf{\alpha}$ and $\mathbf{\beta}$. In summary, probabilistic archetypes lie in the parameter space, whereas classical archetypes lie in the observation space. Thus,  archetypal profiles for binary data are the
probability of a positive response. Details can be found in \cite{Eugster14}.

\subsection{AA and ADA in the functional case}
In Functional Data Analysis (FDA) each datum is a function. Therefore, the sample is a set of functions  $\{x_1(t),...,x_n(t)\}$ with $t \in [a,b]$, i.e. the values of the $m$ variables in the standard multivariate context are replaced by function values with a continuous index $t$. We assume that these functions belong to a Hilbert space, satisfy reasonable smoothness conditions and are square-integrable functions on that interval. Simplistically, the sums are replaced by integrals in the definition of the inner product.

\textcolor{black}{AA and ADA were extended to functional data by \cite{Epifanio2016}. In the functional context, functions from the data set  are approximated by mixtures of archetypal functions.}
In functional archetype analysis (FAA), we seek $k$ archetype functions that approximate the functional data sample by their mixtures. In other words, the objective of FAA is the same as AA, but now both archetypes ($z_j(t)$) and observations  ($x_i(t)$)  are functions. As a consequence, RSS is now calculated with a functional norm  instead of a vector norm. We consider the $L^2$-norm, $\|f\|^2= <f,f> = \int_a^b f(t)^2 dt$. The interpretation of  matrices $\alpha$ and $\beta$ is the same  as in the classical multivariate case.

Analogously, FADA is also a generalization of ADA, where $k$ functional archetypoids, which are functions of the sample, approximate the functions of the sample through the mixtures of these functional archetypoids. Again, vectors are replaced by functions and vector norms by functional norms, and the matrices are interpreted is the same way as before.

To obtain FAA and FADA in a computationally efficient way (\cite{Epifanio2016}), functional data are represented by means of basis functions (see \cite{Ramsay05} for a detailed explanation about smoothing functional data). Let $B_h$  ($h$ = 1, ..., $m$) be the basis functions and $\mathbf{b}_i$ the vector of coefficients of length $m$ such that $x_i(t)$ $\approx$ $\sum_{h=1}^m b_i^h B_h(t)$. Then, RSS is formulated as (see \cite{Epifanio2016} for details):

\begin{equation}\label{RSSfar}
\begin{split}
 RSS = \displaystyle \sum_{i=1}^n \| {x}_i - \sum_{j=1}^k \alpha_{ij} {z}_j\|^2 = \sum_{i=1}^n \| {x}_i - \sum_{j=1}^k \alpha_{ij} \sum_{l=1}^n \beta_{jl} {x}_l\|^2 = 
 \sum_{i=1}^n {\mathbf{a}}'_i \mathbf{W} {\mathbf{a}}_i
 {,}
\end{split}
\end{equation}
where $\mathbf{a'}_i$ = $\mathbf{b'}_i - \sum_{j=1}^k \alpha_{ij} \sum_{l=1}^n \beta_{jl} \mathbf{b'}_l$ and $\mathbf{W}$ is the order $m$ symmetric matrix with the inner products of the pairs of basis functions $w_{m_1,m_2}$ = $\int B_{m_1}B_{m_2}$. If the basis is orthonormal,  for instance the Fourier basis, $\mathbf{W}$ is the order $m$ identity matrix and FAA and FADA can be estimated  using standard AA and ADA with the basis coefficients. If not, $\mathbf{W}$ has to be calculated previously one single time by numerical integration.

\vspace{-0.6cm}

\textcolor{black}{\subsection{Other unsupervised learning techniques} \label{otras}} \textcolor{black}{The following well-known multivariate analysis tools for binary data are used in the comparison.} \textcolor{black}{ We use homogeneity analysis (multiple correspondence analysis) using the R package {\bf homals} \citep{homals} (HOMALS).} \textcolor{black}{HOMALS can be considered as an equivalent to PCA in the case of categorical data. For CLA we use} \textcolor{black}{ Partitioning Around Medoids (PAM) from the R package {\bf cluster} \citep{cluster,Kaufman90}, since it returns representative objects or medoids among the observations of the data set. The pairwise dissimilarities between observations in the data set needed for PAM are computed with the {\it daisy}  function  from the R package {\bf cluster} \citep{cluster}, specifically using Gower's coefficient \citep{gower71similarity} for binary observations. \textcolor{black}{Other popular clustering methods \citep{surveydean} are also used in the comparison: latent class analysis (LCA)  from the R package {\bf poLCA} \citep{polca}, which is a finite mixture model clustering for
categorical data, and classical $k$-means clustering \citep{Lloyd82leastsquares}. It is used in the literature \citep{Henry2015}, despite not being recommended for binary data  \citep{IBM}. For that reason, we also consider PAM, since it is a robustified version of $k$-means \citep{kmeansDouglas} that can be used with distances other than Euclidean, and observations, rather than centroids, serve as the exemplars for each cluster.   }
}





\textcolor{black}{\section{Archetypal analysis for binary data \label{binario}}} 
 Let $\mathbf{X}$ be an $n \times m$ binary matrix with $n$ observations and $m$ variables. The idea behind archetypal analysis is that we can find a set of archetypal patterns, and that data can be expressed as a mixture of those archetypal patterns. In the case of binary data, on the one hand the archetypal patterns should also be binary data, as the population from which data come. For example, if pregnancy was one of the binary variables, 
it would not make sense to consider as an archetypal observation a woman who was pregnant 0.7. In other words, archetypal patterns should be binary  in order to have a clear meaning and not lose their interpretability, which is the cornerstone of archetypal techniques, i.e. they should not be `mythological', but rather something that might be observed. On the other hand, in order to describe data as mixtures, we should assume that observations exist in a vector space, i.e. that observations can be multiplied by scalars (in this case in the interval $\left[ 0, 1 \right]$) and added together.

A solution that meets all these ideas is to apply ADA to $\mathbf{X}$, since the feasible archetypal patterns belong to the observed sample. In fact, ADA was originally created as a response to the problem in which pure non-fictitious patterns were sought \citep{Vinue15}. 

Instead, the archetypes returned by applying AA or PAA do not need to be binary, i.e. they do not need to belong to the feasible set of solutions. In fact, \cite{Eugster14} binarized the archetypes obtained by AA or PAA in  experiments. 
However, using a continuous optimization problem to solve a problem whose feasible solutions are not continuous can fail badly \citep[Ch. 13]{Fletcher2000}. Indeed, there is  no guarantee that this approach will provide a good solution, even by examining all the feasible binary solutions in a certain neighborhood of the continuous solution.

Therefore, we propose to use ADA to handle binary observations.\\

\section{Simulation study}
\label{simulacion} We have carried out a simulation study to assess all the alternatives in a controlled scenario. The design of the experiment has been based on simulation studies that appear in \cite{Vinue15} and \cite{Eugster14}. We generate $k$ = 6 archetypes, ${\bf \zeta}_i$, with $m$ = 10 binary variables by sampling them from a Bernoulli distribution with a probability of success  $p$ = 0.7, $\bf{A}$ = $[{\bf \zeta}_1$, ${\bf \zeta}_2$, ${\bf \zeta}_3$, ${\bf \zeta}_4$, ${{\bf \zeta}}_5$, ${\bf \zeta}_6]$.  Given the archetypes, we  generate $n$ = 100 observations as the binarized  version of $\bf{x}_i$ = $\bf{\tilde{A}}_i \bf{h}_i$ + $\bf{E}_i$, where $\bf{\tilde{A}}_i$ contains the archetypes after adding salt and pepper noise to them, $\bf{h}_i$ is a random vector sampled from a Dirichlet distribution with $\bf{\alpha}$ = (0.8, 0.8, 0.8, 0.8, 0.8, 0.8), and $\bf{E}_i$ is a 10-dimensional random vector of Gaussian white noise with a mean of zero and standard deviation of 0.1. The binarized versions are obtained by replacing all values above 0.5 with 1 and others with 0. The noise density added to $\bf{A}$ is 0.05 (the default value used in MATLAB). With salt and pepper noise, a certain amount of the data is changed to either 0 or 1. To
ensure that $\bf{\tilde{A}}_i$'s are archetypes, we chose $\alpha$ = 0.8, a value near to but less than one. 

We compute PAA, AA and ADA. The archetypes returned by PAA and AA  are binarized for comparison with the true ones, $\bf{A}$. We calculated the Hamming distance (Manhattan distance between binary vectors), which is the same as the misclassification error used with binary images, between each archetypal solution and the true archetypes after permuting the columns of each archetypal solution to match the true archetypes in such a way that the least error with the city block distance is provided.

\begin{figure}[ht]
\centering
\includegraphics[width=0.9\textwidth]{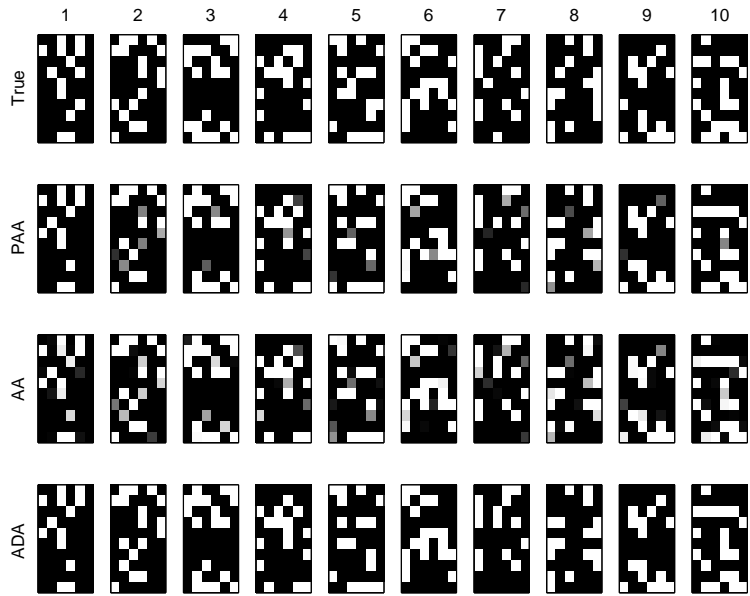}
\caption{\label{fig:simu} Comparison between true archetypes and those returned by PAA, AA and ADA, respectively. The 10 columns represent the first 10 repetitions of the simulation. Black represents 0 and white 1.}
\end{figure}

This was repeated 100 times. The first 10 times are displayed in Figure \ref{fig:simu}. The solutions returned  by all the methods are quite similar to the true archetypes, i.e. the number of errors (a zero in the solution where the true value is 1, or vice versa) is very small. Nevertheless, there are differences between the methods, which are more evident in columns 5 and 6. For columns 5 and 6, the number of errors for PAA is 5 and 5, it is 4 and 2 for AA, but only 2 and 2 for ADA. Table \ref{tab:simu} shows a the summary of the misclassifications. The archetypoids returned by ADA match  the true archetypes better than those returned by AA or PAA, in this order, i.e. ADA provides the smallest mean misclassification error.

\begin{table}[ht]
\caption{\label{tab:simu} Summary of misclassification errors of the archetype profiles for each method over 100 simulations.}
\centering
\begin{tabular}{c|ccc}
Method & PAA & AA &ADA  \\\hline
Mean (Std. dev.) & 4.20 (1.86) &    3.59 (1.99) &   3.19 (1.88)
\end{tabular}
\end{table}

\section{Applications}
\label{aplicaciones} 
\subsection{An initial mathematical skills test for first-year university students} 

\vspace{-0.4cm}
\textcolor{black}{\subsubsection{Data} \label{datos1}}
\textcolor{black}{The first application corresponds to the first point of view of the binary matrix (analysis of the rows). We analyze the data set described by \cite{GregoriOrus}, which was obtained through the application
of a test on the initial mathematics skills  of 690 first-year students of the College
of Technology and Experimental Sciences at  Jaume I University (Spain)  at the beginning of the 2003-04  academic
year. The test consisted of 17 questions corresponding to 21 single items, the answers to which were coded as 0 (incorrect or unanswered) or 1 (correct). The items of the test were selected in order to ascertain some given didactic hypotheses on the didactic discontinuities between  mathematics at  pre-university and university levels.} \textcolor{black}{It is not a test designed to rank the students and return a unique score.} \textcolor{black}{The complete description of the questions can be seen in \cite{GregoriOrus}. With ADA, we could obtain students' skill set profiles. In this way, students can be grouped by their similar mastery of skills.  For instance,
students showing consistently high levels of aptitude may be selected for an advanced
class or students with similar difficulties
could receive extra instruction as a group and also teaching strategies  could also be adapted to suit their level. A classical way to group student skill set profiles is by using a clustering method, as carried out by \cite{DeanNuget}, but in terms of human interpretability, the central points returned by clustering tools do not seem as favorable as the extreme points returned by ADA. {Results from different exploratory tools are compared.}}

\textcolor{black}{\subsubsection{Results and discussion}}
We would like to estimate the skill set profiles hidden in the data set.  
In other words, we would like to discover the data structure. Our intuition tells us that skill sets vary continuously across students, i.e. we do not expect there to be clearly differentiated (separate)   groups of students with different abilities. Even so, CLA has been used to generate groups of students with similar skill set profiles \citep{Chiu2009,DeanNuget}. 
Here, we are going to consider the raw binary data and let the data speak for themselves, as ADA is a data-driven method. We compare the ADA solution with others from well-established unsupervised techniques \textcolor{black}{introduced in Section \ref{otras}} to highlight the information about the
quality understanding of data provided by ADA.

\begin{figure}[!htbp] 
  \centering
  \subfloat{\includegraphics[width=0.4\textwidth]{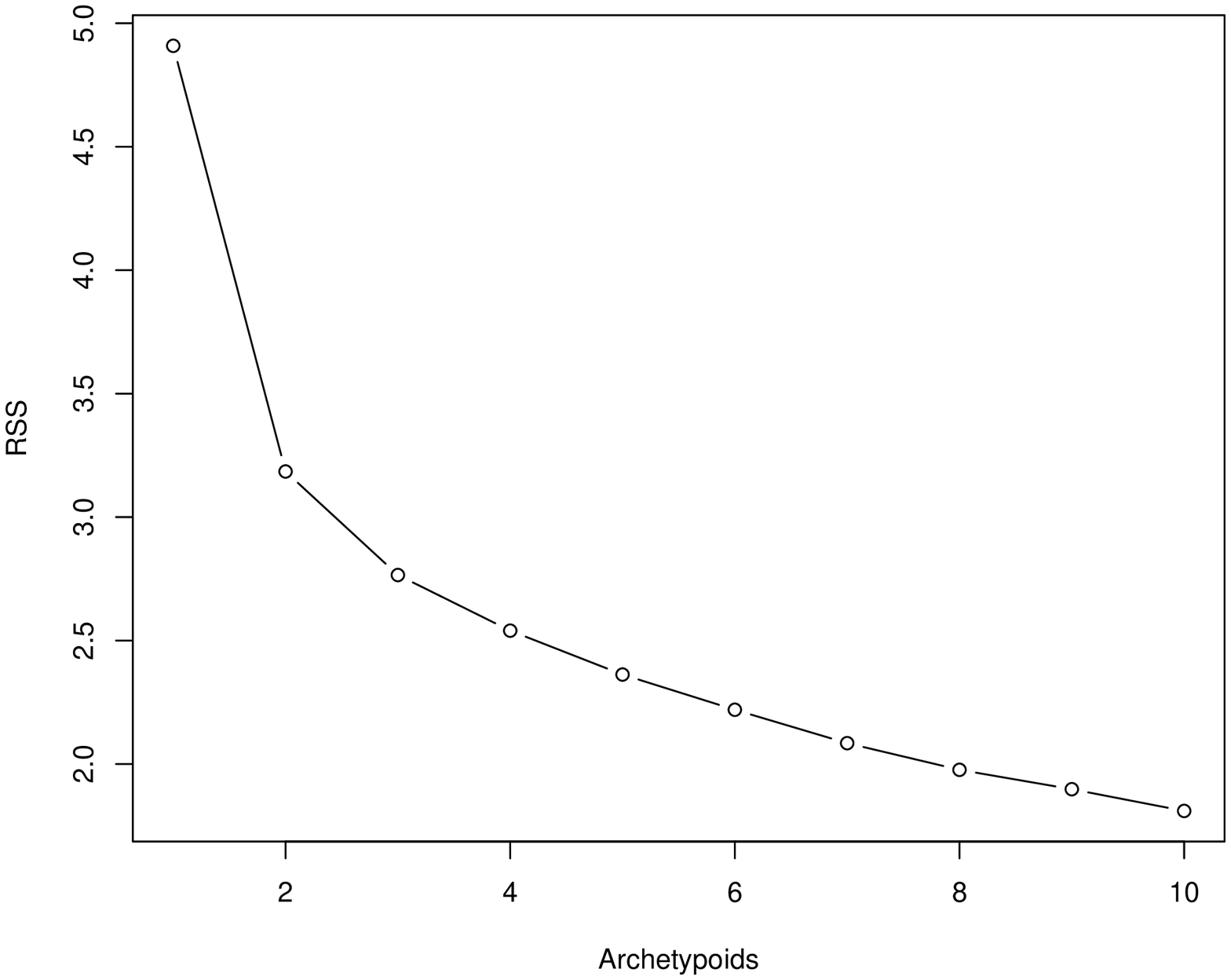}}
  \subfloat{\includegraphics[width=0.4\textwidth]{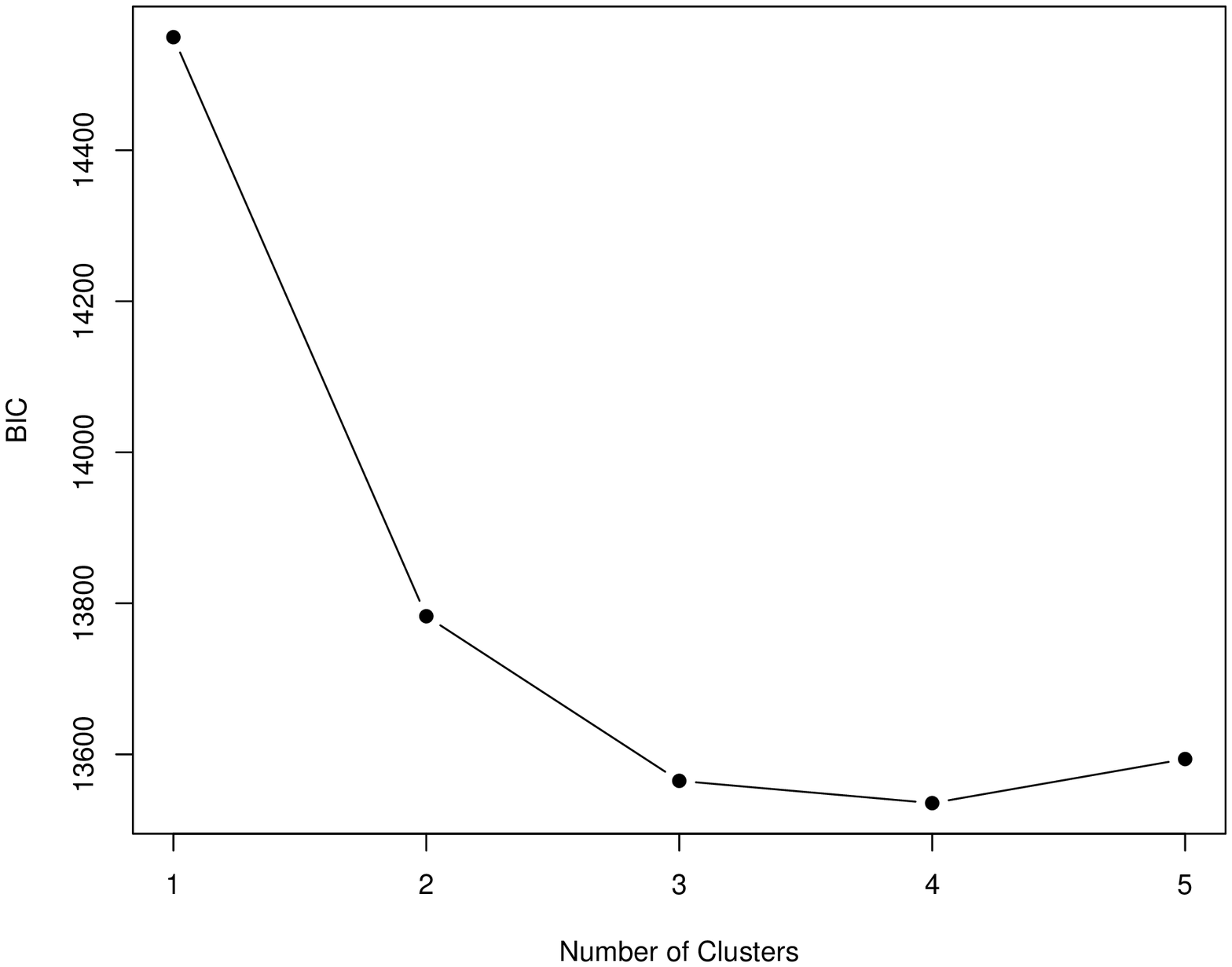}}
  \caption{\textcolor{black}{Initial mathematical skills test data:  Screeplot of ADA (left-hand panel); screeplot of LCA (right-hand panel).} \label{screeplot} } 
\end{figure}

\begin{figure}[!hbt]
\centering
\includegraphics[width=0.4\textwidth]{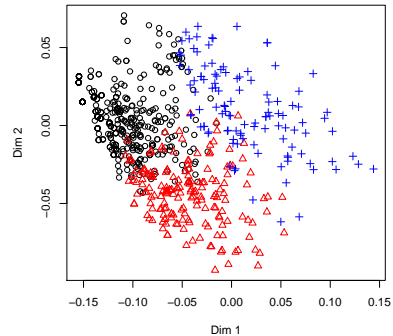}
\caption{\label{hplot} H-plot of dissimilarities for the initial mathematical skills test data. We perform PAM. The black circles represent data points assigned to the first cluster, the red triangles to the second cluster and the blue crosses to the third cluster.}
\end{figure}

For the sake of
brevity and as an illustrative example, we examine the results of $k$ = 3. The RSS elbow for ADA \textcolor{black}{and the Bayesian  Information Criterion (BIC) elbow for LCA are} found at $k$ = 3 (see Figure \ref{screeplot}). \textcolor{black}{According to the silhouette coefficient \textcolor{black}{(a method of interpretation and validation of consistency within clusters of data, see \cite{Kaufman90} for details)}, the optimal number of clusters are $k$ = 2 and $k$ = 3 for PAM. However, }
the highest value of the silhouette coefficient  is 0.22 (for $k$ = 2 and $k$ = 3 clusters), which means that no substantial cluster structure was found, as we predicted. We perform an h-plot (a multidimensional scaling method that is particularly suited for representing non-Euclidean dissimilarities, see \cite{Epi2013} for details) on the dissimilarities used by PAM to graphically summarize the data set  and to visualize the obtained clusters \textcolor{black}{by PAM} in two dimensions (see Figure \ref{hplot}). Effectively, separate clusters do not seem to exist.

%


\begin{figure}[!hbt]
\centering
  \subfloat{\includegraphics[width=0.4\textwidth]{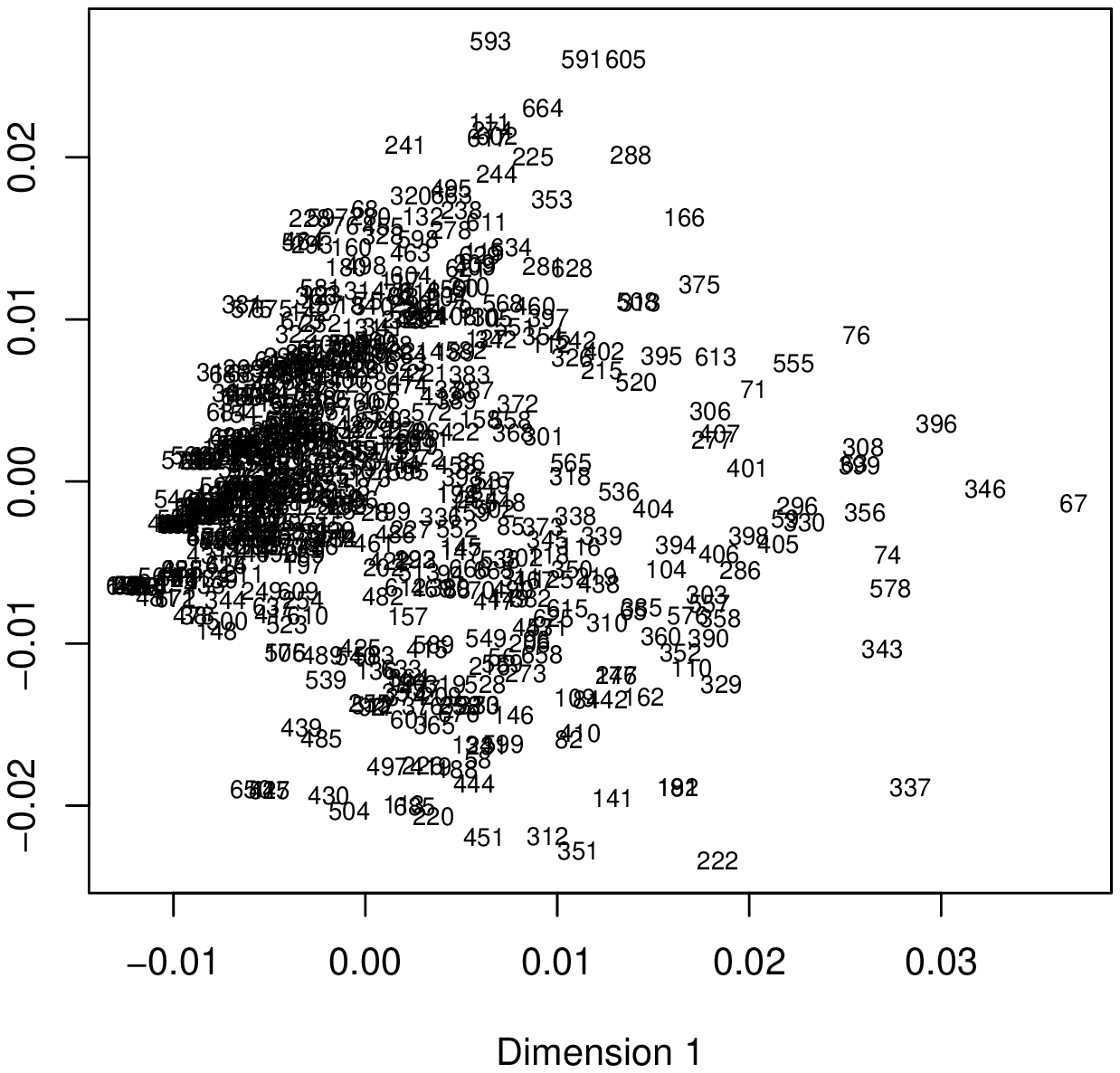}}
  \subfloat{\includegraphics[width=0.4\textwidth]{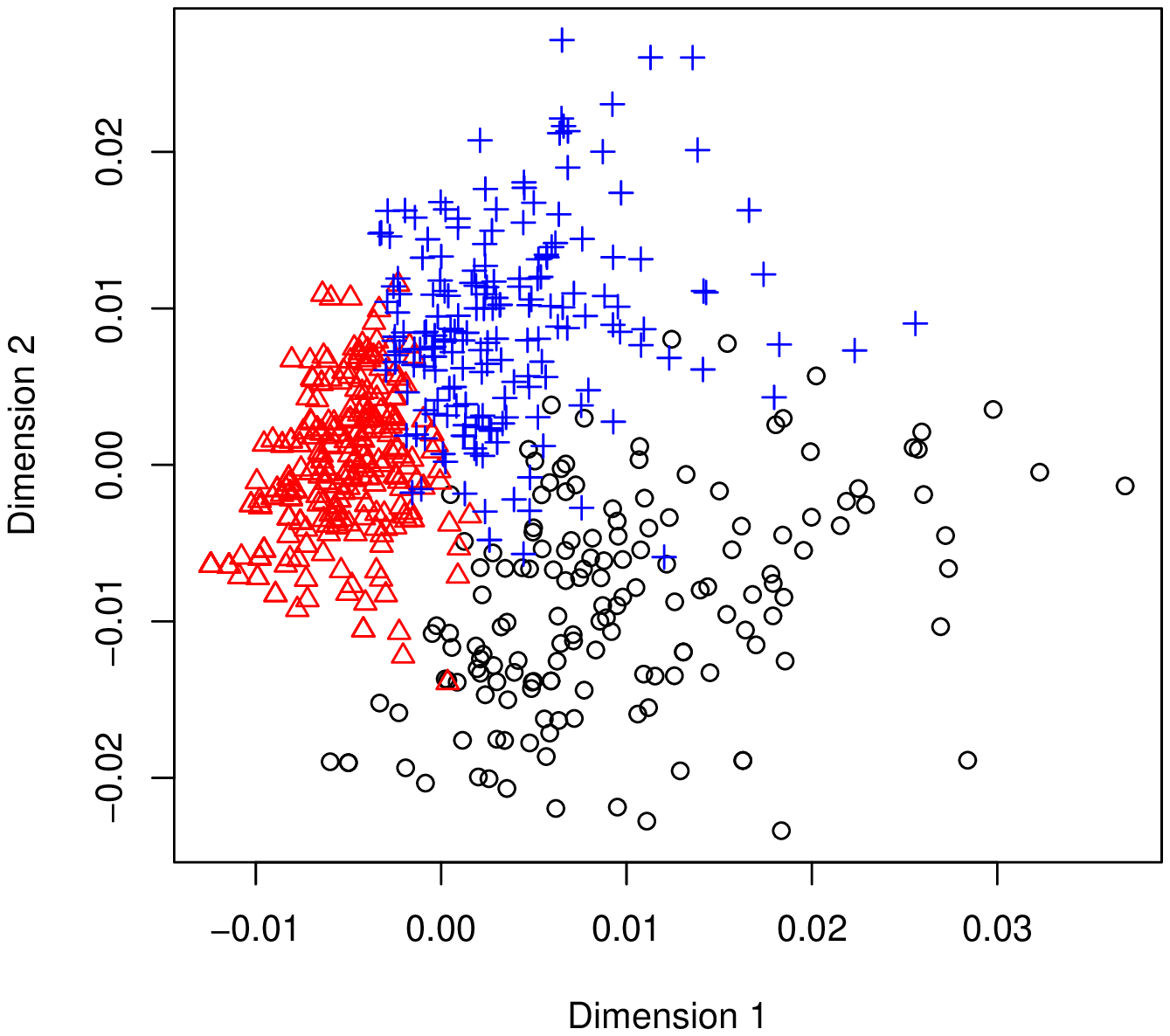}}
\caption{\label{homalsfig1} HOMALS of the initial mathematical skills test data. Plot of students' scores. The numbers indicate the code of each of the 690 students \textcolor{black}{(left-hand panel). We perform LCA. The black circles represent data points assigned to the first cluster, the red triangles to the second cluster and the blue crosses to the third cluster (right-hand panel).} }
\end{figure}

\begin{figure}[!hbt]
\centering
 \includegraphics[width=0.5\textwidth]{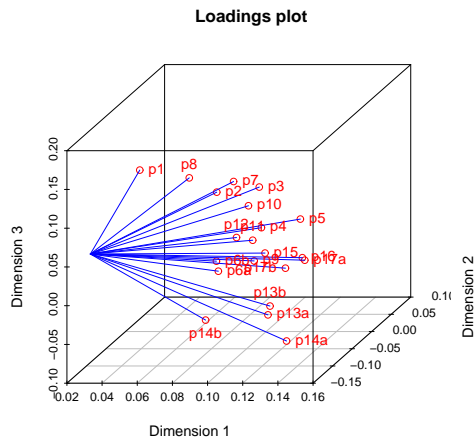}
\caption{\label{homalsfig2} HOMALS of the initial mathematical skills test data. Loadings plot.}
\end{figure}

This is also corroborated by Figure \ref{homalsfig1}, where the students' scores from HOMALS are plotted in two dimensions.  As regards the interpretation of the dimensions of HOMALS, the loadings are displayed in Figure \ref{homalsfig2} and Table \ref{loadingstable} shows their exact values, together with the number of correct answers. As also happens with PCA, their interpretation is not always easy and immediate. For the first dimension, all the coefficients are positive (as a measure of size), which can indicate a kind of sum score. The highest coefficients more or less correspond to the last questions of the test, which fewer students answered correctly. The second dimension compares, above all, questions 4, 5, 6a and 6b (with high positive coefficients) with 13a and 13b (with low negative values), while in the third dimension, questions 1, 3, 7, 8 and 10 (with high positive coefficients)  are compared with  14a and 14b (with low negative values). However, we do know how the meaning of these contradistinctions is interpreted.

 \textcolor{black}{LCA returns the conditional item response probabilities by outcome variable for each class. Table \ref{perfiles} lists these probabilities for correct answer. The predicted classes for each student are shown in Figure \ref{homalsfig1}, since the profiles of cluster 1 and 3 are mainly differentiated in questions 4, 5, 13a and 13b, which are the most relevant variables of dimension 2 of HOMALS.}

\begin{table}[!hbt]
\begin{center}
\caption{Number of correct answers and loadings of the first three dimensions by HOMALS for the initial mathematical skills test data. \label{loadingstable}}
\begin{scriptsize}
\begin{tabular}{ccccc}
\hline
Question & No. correct answers & D1& D2& D3\\
\hline
1  & 	621	&	0.02862783& 	-0.0023240445&  	0.109355256\\
\hline
2  & 	589	&	0.05796987  &	0.0626839294  &	0.052003283\\
\hline
3   &	301	&	0.09098387 	&0.0229084678  &	0.076212734\\
\hline
4   &	233	&	0.07597107  &	0.0959592091 &	-0.008635664\\
\hline
5   &	253	&	0.09922374  &	0.0910173149&  	0.004650411\\
\hline
6a 	&231	&	0.05413846&  	0.0839718688 &	-0.059468794\\
\hline
6b &	105	&	0.05230047 & 	0.0873491670 &	-0.048042860\\
\hline
7 &  	270	&	0.07408320  &	0.0332456303 & 	0.078814530\\
\hline
8  & 	140	&	0.06009601 &	-0.0172906782 & 	0.105955795\\
\hline
9   &	109	&	0.07749049  &	0.0705908727 &	-0.040283362\\
\hline
10 &	202	&	0.09540734 &	-0.0246907330 & 	0.073252540\\
\hline
11  &	71	&	0.07484609 & 	0.0786972991& 	-0.017170762\\
\hline
12&  	329	&	0.08006423 & 	0.0138046564  &	0.015179516\\
\hline
13a	&177	&	0.12934508& 	-0.1274837375 &	-0.021853917\\
\hline
13b &	132		&0.12952677 &	-0.1231243539 &	-0.012406654\\
\hline
14a 	&114	&	0.11951449 &	-0.0355861709 	&-0.096428155\\
\hline
14b &	22	&	0.07564891& 	-0.0454400932 &	-0.065063675\\
\hline
15 & 	183	&	0.10748607 &	-0.0366512703 & 	0.017544829\\
\hline
16 & 	236	&	0.12062274 	&-0.0001786048& 	-0.004647963\\
\hline
17a &	47	&	0.12115852  &	0.0035393336 &	-0.009522369\\
\hline
17b &	62		&0.10884146 &	0.0095484411 &	-0.022486862\\
\hline
\end{tabular}
\end{scriptsize}
\end{center}
\end{table}

Table \ref{perfiles} \textcolor{black}{also} lists the profiles of the medoids\textcolor{black}{, centroids of $k$-means} and the archetypal profiles for AA, PAA and ADA. For medoids and archetypoids, the code of the corresponding observation is also displayed. To facilitate the analysis we also show the binarized profiles of AA and PAA\textcolor{black}{, referred as BAA and BPAA, respectively}.

As a simple summary of the profiles, we compute the percentage of correct answers for each profile. For PAM, the percentages are 9.5\%, 33.3\% and 57.1\%; \textcolor{black}{for binarized LCA, 38.1\%, 9.5\% and 33.3\%; for binarized $k$-means, 38.1\%, 9.5\% and 42.9\%;} for BAA, 9.5\%, 47.6\% and 61.9\%; for BPAA,  9.5\%,   42.9\% and   57.1\%; and for ADA, 57.1\%, 52.4\% and  9.5\%, respectively. Note that the median of the percentage of correct answers in the data set is 28.6\% (the minimum is 0, the first quartile is  19.1\%, the third quartile is 38.1\%, while the maximum is 95.2\%).

One profile is repeated in all the methods, a student who only answers  questions 1 and 2 correctly, i.e. a student with a serious lack of competence. We therefore concentrate the analysis on the other two profiles for each method.

In contrast with the third archetypoid, i.e. the student with very poor skills, the first and second archetypoids correspond to students with  very high percentages of correct answers. In fact, the first archetypoid corresponds to the 92nd percentile of the data set, while the second archetypoid corresponds to the 88th percentile. Nevertheless, both profiles are quite different. In fact, the Hamming distance between archetypoids 1 and 2 is 13, which means that although they answered  a lot of items correctly, these correctly answered items do not coincide. In other words, archetypoids 1 and 2 are somehow complementary. Both answered items 1, 2, 3, 12 and 16 correctly, which were among the most correctly answered items. Neither of them answered items 11, 14b and 17a correctly,  which were among the least correctly answered items. On the one hand, the items that archetypoid 1 answered correctly, but archetypoid 2 did not are  8, 10, 13a, 13b, 14a, 15 and 17b. These items are about nonlinear systems and linear functions. On the other hand, the items that archetypoid 2 answered correctly, but archetypoid 1 did not are  4, 5, 6a, 6b, 7 and 9. These items are about the calculation of derivatives and integrals and algebraic interpretation. The skills of these archetypoids are clear and different to each other.

\begin{figure}[!hbt]
\centering
 \includegraphics[width=0.5\textwidth]{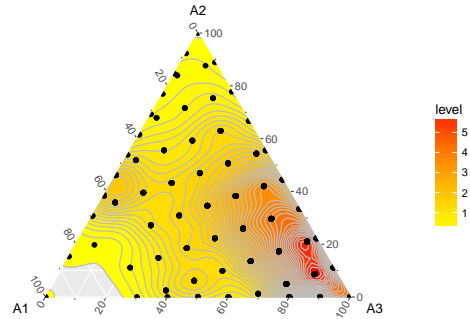}
\caption{\label{ternario} Ternary plot of {\bf $\alpha$}s of ADA together with a plot density estimate for the initial mathematical skills test data.}
\end{figure}

We can use the alpha values for each of
the students to learn about their relationship to the archetypoid profiles. 
The ternary plot in Figure \ref{ternario} displays the alpha values
that provide further insight into the data structure. Note that the majority of the data is concentrated around archetypoid 1, i.e. the one with very poor skills. If we wanted to form three groups using the alpha values, we could assign each student to the group in which their  corresponding alpha is the maximum\textcolor{black}{, as we did in Figure \ref{toy} (d)}. In this way, the number of students similar to archetypoid 1 is 113, to archetypoid 2 it is 110 and to archetypoid 3 it is 467.

\begin{table}[!hbt]
\setlength{\tabcolsep}{1pt}
\begin{center}
\footnotesize
\caption{Profiles for the initial mathematical skills test data, for PAM, \textcolor{black}{LCA, $k$-means ($k$-M)}, AA (and binarized, BAA), PAA (and binarized, BPAA) and ADA, with $k$ = 3. The numbers in brackets for PAM and ADA indicate the code of the representative student \label{perfiles} }
\begin{tabular}{lccccccccccccccccccccc}
Methods & 1 & 2& 3& 4& 5 & 6a& 6b& 7& 8& 9& 10& 11& 12 &13a& 13b& 14a& 14b& 15& 16& 17a& 17b\\
\hline
PAM (661)  &1 &  1 &  0 &  0 &  0   & 0 &    0  &  0& 0  & 0& 0  &  0&  0 &  0 &   0  &  0 &   0&   0 &  0 &  0 &  0\\
\hline
PAM (586) &	1 &  1 &  1 &  1&   1&    0 &    0&    1 &  0 &  0 &   0&    0 &   1 &  0 &  0 &  0&  0&  0&  0&  0 & 0\\
\hline
PAM (162) & 1 &  1 &  1 &  0 &  1 &   0 &    0 &   1  & 0 &  0 &   1 &   0 &   1  &  1 &  1 &  1 &  0 & 1 & 1 & 0 &  0\\
\hline
\textcolor{black}{LCA 1} &\textcolor{black}{0.93}&\textcolor{black}{0.88}&\textcolor{black}{0.57}&\textcolor{black}{0.36}&\textcolor{black}{0.48}&\textcolor{black}{0.37}&\textcolor{black}{0.17}&\textcolor{black}{0.49}&\textcolor{black}{0.32}&\textcolor{black}{0.20}&\textcolor{black}{0.51}&\textcolor{black}{0.14}&\textcolor{black}{0.62}&\textcolor{black}{1.00}&\textcolor{black}{0.85}&\textcolor{black}{0.43}&\textcolor{black}{0.10}&\textcolor{black}{0.50}&\textcolor{black}{0.55}&\textcolor{black}{0.18}&\textcolor{black}{0.19} \\ \hline
\textcolor{black}{LCA 2} &\textcolor{black}{0.88}&\textcolor{black}{0.79}&\textcolor{black}{0.28}&\textcolor{black}{0.17}&\textcolor{black}{0.15}&\textcolor{black}{0.24}&\textcolor{black}{0.07}&\textcolor{black}{0.30}&\textcolor{black}{0.14}&\textcolor{black}{0.03}&\textcolor{black}{0.15}&\textcolor{black}{0.02}&\textcolor{black}{0.32}&\textcolor{black}{0.05}&\textcolor{black}{0}&\textcolor{black}{0.03}&\textcolor{black}{0}&\textcolor{black}{0.13}&\textcolor{black}{0.14}&\textcolor{black}{0}&\textcolor{black}{0} \\ \hline
\textcolor{black}{LCA 3} &\textcolor{black}{ 0.91}&\textcolor{black}{0.94}&\textcolor{black}{0.60}&\textcolor{black}{0.62}&\textcolor{black}{0.65}&\textcolor{black}{0.48}&\textcolor{black}{0.27}&\textcolor{black}{0.47}&\textcolor{black}{0.23}&\textcolor{black}{0.35}&\textcolor{black}{0.36}&\textcolor{black}{0.23}&\textcolor{black}{0.63}&\textcolor{black}{0.04}&\textcolor{black}{0}&\textcolor{black}{0.20}&\textcolor{black}{0.03}&\textcolor{black}{0.31}&\textcolor{black}{0.53}&\textcolor{black}{0.09}&\textcolor{black}{0.17}\\ \hline
\textcolor{black}{$k$-M 1}&\textcolor{black}{0.91}&\textcolor{black}{0.95}&\textcolor{black}{0.64}&\textcolor{black}{0.70}&\textcolor{black}{0.74}&\textcolor{black}{0.43}&\textcolor{black}{0.25}&\textcolor{black}{0.53}&\textcolor{black}{0.24}&\textcolor{black}{0.32}&\textcolor{black}{0.38}&\textcolor{black}{0.22}&\textcolor{black}{0.63}&\textcolor{black}{0.05}&\textcolor{black}{0.00}&\textcolor{black}{0.19}&\textcolor{black}{0.02}&\textcolor{black}{0.33}&\textcolor{black}{0.52}&\textcolor{black}{0.10}&\textcolor{black}{0.15}\\\hline
\textcolor{black}{$k$-M 2}&\textcolor{black}{0.88}&\textcolor{black}{0.78}&\textcolor{black}{0.26}&\textcolor{black}{0.13}&\textcolor{black}{0.11}&\textcolor{black}{0.27}&\textcolor{black}{0.09}&\textcolor{black}{0.27}&\textcolor{black}{0.13}&\textcolor{black}{0.05}&\textcolor{black}{0.14}&\textcolor{black}{0.02}&\textcolor{black}{0.32}&\textcolor{black}{0.06}&\textcolor{black}{0.01}&\textcolor{black}{0.03}&\textcolor{black}{0.01}&\textcolor{black}{0.12}&\textcolor{black}{0.15}&\textcolor{black}{0.01}&\textcolor{black}{0.01}\\\hline
\textcolor{black}{$k$-M 3}&\textcolor{black}{0.93}&\textcolor{black}{0.91}&\textcolor{black}{0.59}&\textcolor{black}{0.34}&\textcolor{black}{0.49}&\textcolor{black}{0.37}&\textcolor{black}{0.17}&\textcolor{black}{0.50}&\textcolor{black}{0.33}&\textcolor{black}{0.20}&\textcolor{black}{0.54}&\textcolor{black}{0.14}&\textcolor{black}{0.64}&\textcolor{black}{1}&\textcolor{black}{0.88}&\textcolor{black}{0.46}&\textcolor{black}{0.10}&\textcolor{black}{0.52}&\textcolor{black}{0.58}&\textcolor{black}{0.18}&\textcolor{black}{0.19}\\\hline
AA 1& 0.85&0.68&0.02&0&0&0.05&0&0.04&0.05&0&0.01&0&0.16&0&0&0&0&0.01&0&0&0 \\ \hline
AA 2& 0.90&1&0.87&1&1&1&0.63&0.82&0.19&0.52&0.24&0.38&0.65&0&0&0.16&0&0.07&0.43&0.09&0.15\\ \hline
AA 3& 1&1&0.89&0.32&0.53&0.19&0.06&0.71&0.58&0.26&1&0.17&1&1&1&0.67&0.18&1&1&0.36&0.37\\ \hline
BAA 1& 1&1&0&0&0& 0& 0&0&0&0& 0& 0& 0&0&0&0&0& 0& 0&0&0\\ \hline
BAA 2& 1&1&1&1&1& 1& 1&1&0&1& 0& 0& 1&0&0&0&0& 0& 0&0&0\\ \hline
BAA 3& 1&1&1&0&1& 0& 0&1&1&0& 1& 0& 1&1&1&1&0& 1& 1&0&0\\ \hline
PAA 1 &        0.86    &  0.72    &  0.13   &  0   &  0    &  0   &  0   &  0.12   &  0.07   &  0   &  0   &   0  &    0.23  &    0   &      0  &      0    &   0      &   0   &   0     &  0     &   0\\
\hline
PAA 2 &        0.90    &  1    &  0.78  &   1 &    1   &   1  &   0.61   &  0.73  &   0.27  &   0.40   &  0.38  &    0.31   &   0.66    &  0     &    0     &   0    &   0     &    0    &  0.43  &     0   &     0\\
\hline
PAA 3 &        0.99  &    1   &   0.82   &  0.36  &   0.57   &   0.25   &  0  &   0.66 &    0.44   &  0.27 &    0.86    &  0.12  &    0.85   &   1   &      1    &    0.73   &    0.15   &      1  &    1   &    0.32   &     0.42\\ \hline
BPAA 1 &  1 &  1 &   0  &  0   & 0 &    0 &   0   & 0 &   0  & 0    & 0    & 0    & 0    & 0   &    0 &      0      & 0 &     0       & 0    &  0    &  0\\
\hline
BPAA 2 &  1 &  1   & 1 &   1    &1  &   1    &1    &1&    0 &  0     &0&     0 &    1    & 0   &    0 &      0    &   0     & 0      & 0 &     0  &    0\\
\hline
BPAA 3 &  1 &  1   & 1   & 0  &  1  &   0   & 0  &  1 &   0 &  0   &  1    & 0    & 1  &   1     &  1    &   1   &    0  &    1       &1  &    0   &   0\\ \hline
ADA (182) & 1 & 1 & 1 & 0&  0 & 0 &   0 & 0 &  1 & 0&  1 &  0 &  1  & 1&    1  &  1&    0  &  1 &   1&   0  & 1 \\
\hline
ADA (274) & 1&  1&  1&  1&  1 & 1 &   1 & 1 &  0&  1&  0 &  0  & 1 &  0 &   0   & 0&    0  &  0   & 1 &  0  & 0 \\
\hline
ADA (1)  &  1 & 1 & 0 & 0 & 0&  0 &   0&  0 &  0 & 0&  0 &  0 &  0 &  0 &   0  &  0  &  0  &  0 &   0&   0 &  0\\
\hline
\end{tabular}
\end{center}
\end{table}

The profiles of medoids 2 and 3 are not as complementary as the previous archetypoids. In fact, medoid 2 corresponds to the 56th percentile, while medoid 3 corresponds  to the 92nd percentile. In this case, the percentage of correct answers for  medoid 2 is not high. The Hamming distance between the two medoids is only 7. On the one hand, both answered items 1, 2, 3, 5, 7 and 12 correctly, which are the most correctly  answered items.  On the other hand, both failed  items 6a, 6b, 8, 9, 11, 14b, 17a and 17b, many more items than in the case of ADA. The only item that medoid 2 answered correctly but medoid 3 did not is item 4. The items that medoid 3 answered correctly but medoid 2 did not are 10, 13a, 13b, 14a, 15 and 16. It seems as if the cluster definition was guided by the number of correct answers rather than by the kind of item  answered correctly. This is the reason why PAM selects medoid 2 in the middle of the data cloud. PAM, and usual clustering methods, tries to cover the set in such a way that every point is near to one medoid or one cluster
center. The number of students belonging to each cluster is 398, 179 and 113, respectively. Note that the size of the cluster of  students with poor skills is smaller than  in the case of ADA, because some of those students are assigned to the cluster of medoid 2.

\textcolor{black}{The binarized profile of LCA 1, corresponding to the 75th percentile, is similar to medoid 3, but with a lower number of correct answers (5, 7, 14a and 15), while the binarized profile of LCA 3, corresponding to the 56th percentile, is similar to medoid 2, only differentiated by two items (7 and 16). Therefore, they are even less complementary than the previous medoids. The Hamming distance between both LCA-profiles is only 5. The number of students belonging to each cluster is 155, 352 and 183, respectively. Note that the size of the cluster of  students with poor skills is smaller than in the case of PAM.}

\textcolor{black}{The binarized profile of the first centroid of $k$-means, corresponding to the 75th percentile, is similar to medoid 2, only differentiated by item 16, while the binarized profile of the third centroid, corresponding to the 82nd percentile, is similar to medoid 3, but with a lower number of correct items (5, 7 and 14a). The Hamming distance between both centroids is 7. The level of complementarity between both centroids is similar to that of the medoids of PAM, but the number of correct answers of medoid 3 is higher than binarized centroid 3.   The number of students belonging to each cluster is 196, 349 and 145, respectively. Note that the size of the cluster of  students with poor skills is smaller than in the case of PAM, but larger than in the case of PAM for  cluster 3, which in both clustering methods corresponds to the students with more correct answers.}

\textcolor{black}{In the clustering methods, the profiles of each cluster are not as extreme as archetypoids. Archetypoids are also more complementary, which makes it clearer to establish which kinds of features distinguish one group from another. } \textcolor{black}{Remember also that clustering is limited to assign each student to a group but alpha values of ADA allows to know the composition, i.e. ADA returns a richer information.}

The profiles of BAA2 and BAA3 and BPAA2 and BPAA3 are quite similar to the profiles of archetypoid 2 and 1, respectively, but with slight differences. The percentiles corresponding to correctly  answered items are also high, although for one of the archetypes not as high as for archetypoids. The percentiles are the 82nd and 94th for BAA2 and BAA3, and the 75th and 92nd for BPAA2 and BPAA3, respectively. Therefore, the archetypoids are more extreme than the binarized archetypes of AA and PAA. Although the profiles for BAA and BPAA  are also complementary, they are not as complementary as the two archetypoids. The Hamming distance between BAA2 and BAA3 is 11, and 9 between BPAA2 and BPAA3. Archetypoids therefore manage  to find more complementary profiles.

\subsection{An \textcolor{black}{American College
Testing} (ACT) Mathematics Test} 
\vspace{-0.4cm}

\textcolor{black}{\subsubsection{Data}}
\textcolor{black}{This application \textcolor{black}{ corresponds to the second point of view of the binary matrix (analysis of the columns).}  We use the same data and approach followed by 
 \citet[Ch. 9]{Ramsay02} and \cite{10.2307/3648139}, although another strategy could be considered \citep{doi:10.3102/1076998616680841}. The data used are the 0/1 (incorrect/correct) responses of 2115 males from  administration of a version of the ACT Program 60-item Mathematics Test.} \textcolor{black}{Unlike the test introduced in Section \ref{datos1}, the objective of the test is to relate a student's ACT score with probability of him or her earning a college degree, i.e. to rank students. It seeks that the difficulty of questions increases as you get to higher question numbers.}
 
 \textcolor{black}{ Although this binary matrix does not seem curvaceous at first sight, by making the  simplifying assumption that the probabilities $P_{ih}$ (probability that
examinee $h$ gets item $i$ right) vary in a smooth one-dimensional way across examinees, we can estimate
the ability space curve that this assumption implies. Then, we can work with item response functions \textcolor{black}{(IRFs)} $P_i(\theta)$ as functional data \citep{Ramsay05}, where $\theta$ is the charting variable that measure\textcolor{black}{s} out positions along the ability space curve. Or rather, we can work with 
 the log odds-ratio functions
$W_i(\theta)$, since these transformations of the item response functions have the
unconstrained variation that we are used to seeing in directly observed
curves.  \citet[Ch. 9]{Ramsay02} and \cite{10.2307/3648139} used functional PCA (FPCA) to study variations among these functions. Instead, we propose to use functional ADA (FADA), which reveals very interesting patterns that were not discovered with FPCA.  }

\textcolor{black}{
Note that in the literature, we find other terms for IRFs, such as option characteristic curves, category characteristic curves, operating characteristic curves, category response functions, item category response functions or option response functions \citep{KernSmoothIRT}.}

\begin{figure}[ht]
\centering
\includegraphics[width=0.9\textwidth]{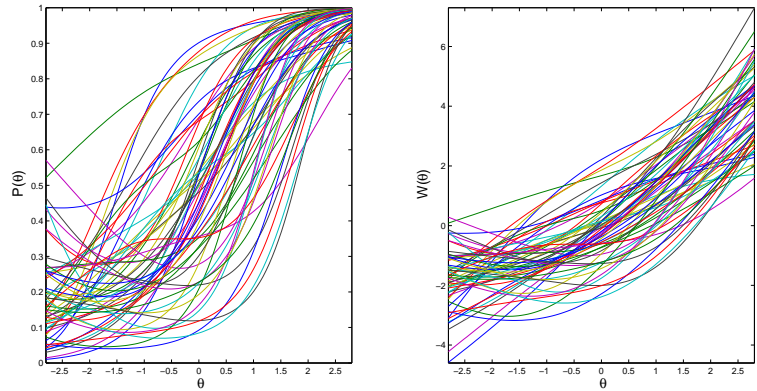}
\caption{\label{PWfig} Estimated IRFs (left-hand panel) and log odds-ratio functions (right-hand panel) for the ACT math exam estimated from the male data.}
\end{figure}

\textcolor{black}{\subsubsection{Results and discussion}}
As mentioned previously, we used the same data and approach followed by 
 \citet[Ch. 9]{Ramsay02} and \cite{10.2307/3648139} to estimate IRFs, $P_i(\theta)$, and their logit functions, $W_i(\theta)$ = $log (P_i(\theta) / (1- P_i(\theta)))$. In particular, a penalized EM algorithm was used and functions were expanded by terms of 11 B-spline basis functions using equally spaced knots. Figure \ref{PWfig} displays the estimated IRFs, $exp(W_i(\theta))/(1 + exp(W_i(\theta)))$, and their log odds-ratio functions $W_i(\theta)$  for the 60 items. As expected, this kind of graphs with superimposed curves is largely uninformative and aesthetically unappealing \citep{JonesRice}.

 To explore a set of curves \cite{JonesRice} proposed the use of functions with extreme principal component scores.  This could be viewed as finding the archetypoid functions. Nevertheless, the aim of PCA is not  
  to recover extreme patterns. In fact, curves with extreme PCA scores do not necessarily correspond to archetypal observations. This is discussed in \cite{Cutler1994} and shown in \cite{EpiVinAle} through an example where  archetypes could not be restored with PCA, even if all the components
had been considered. Not only that, \cite{STONE1996110} also showed that AA may be more appropriate than PCA when the data do not have elliptical distributions.

\begin{figure}[ht]
\centering
\includegraphics[width=0.9\textwidth]{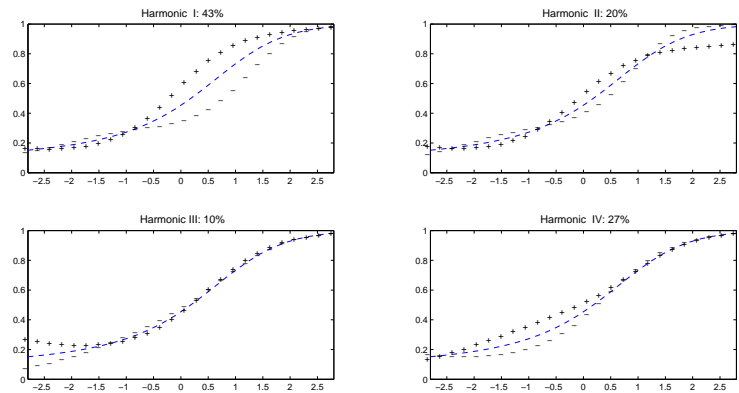}
\caption{\label{FPCAfig} The first four functional PCs in IRFs
after VARIMAX rotation. Plus (negative) signs indicate the effect of adding (subtracting) a multiple of a component
to the mean function. The mean function is the dashed line.}
\end{figure}

In order to show the advantages of ADA over PCA, we compute FPCA and FADA for $W(\theta)$, since they are unconstrained\textcolor{black}{, therefore making them more appropriate for PCA application than the bounded $P_i(\theta)$. This is not a problem with FADA as it works with convex combinations}. Figure \ref{FPCAfig} displays the first four PCs after a varimax rotation having been back-transformed to their probability counterparts, as performed by \citet[Ch. 9]{Ramsay02} and \cite{10.2307/3648139}. We base the interpretation of each PC on the detailed description carried out by \citet[Ch. 9]{Ramsay02}.

\begin{figure}[ht]
\centering
\includegraphics[width=0.9\textwidth]{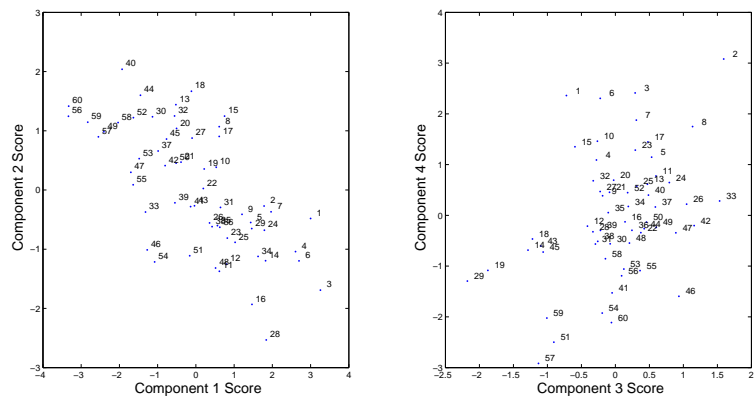}
\caption{\label{pcascores} Bivariate plots of principal component scores of IRFs.  PC1 versus PC2 (left-hand panel); PC3 versus PC4 (right-hand panel).}
\end{figure}

The percentage of total variation explained by those four components is nearly 100\%, while the percentage explained by each component is reported in Figure \ref{FPCAfig}. The first component concentrates on the middle part of the ability range, in such a way that an item with a high (low) score in that component has a higher (lower) slope than the mean from approximately 0 to 2, i.e. it quantifies a discriminability trade-off between
average students and those with rather high abilities. Analogously, the fourth component quantifies a discriminability trade-off between
average examinees and those with rather low abilities. On the contrary, the second component concentrates on the upper end of the ability range. \textcolor{black}{As \cite{10.2307/3648139} explained,  the 3PL model is not well suited to modeling this type of variation.} An item with a low score on this component is
good at sorting out very high ability students from others of
moderately high ability, whereas if the score for this item is high, it will discriminate well
among most of the population but will be found to be of approximately  equal
difficulty by all the very good students. Nevertheless, conclusions on the extreme part of the ability range should be made with caution, since the estimation is carried out  using a
relatively small numbers of students. The third component also accounts
for variation in the characteristics of test items in the extreme ability range, but now in low ability ranges. PC scores for these four components can be seen in Figure \ref{pcascores}. Note that to evaluate the 4 PC scores simultaneously and combine them to give an idea about each item, it is not easily comprehensible or human-readable.

\begin{figure}[ht]
\centering
\includegraphics[width=0.6\textwidth]{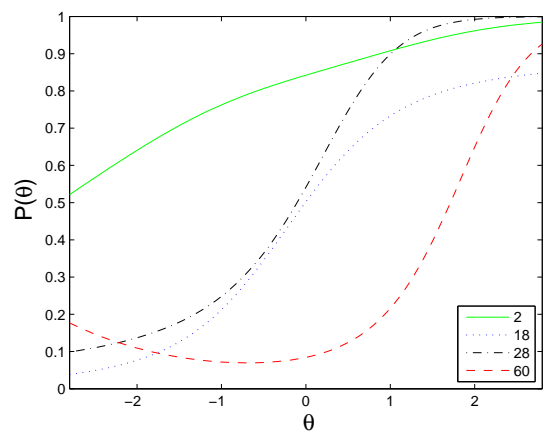}
\caption{\label{FADAfig} ACT data: The four IRF archetypoids are items 2, 18, 28 and 60. See the legend inside the plot.}
\end{figure}

Figure \ref{FADAfig} displays the archetypoids for $k$ = 4  explaining 97\% of the variability, which is nearly as high as FPCA. The archetypoids are  items 2, 18, 28 and 60. These four items describe the extreme patterns found in the sample. Item 2 has very high scores in PC 3 and PC 4, high scores in PC 1 and a score of nearly zero for PC 2. Its IRF is quite flat with a very slight slope, it seems to be a very easy  item, with high  probabilities of success  throughout the ability range.   The other archetypoids discriminate better between low and high ability students but in very different ways. Item 18 has a very high score for PC 2 and a negative score  for PC 3, but nearly zero for PC 1 and PC 4.  It is an item that is quite difficult even for the students in the very high ability range. The IRF of item 28 is quite similar to that of item 18 for the low ability range until  $\theta$  0, but its slope for the high ability range is higher, and the  probabilities of success are higher than 0.9 for $\theta$s higher than 1. On the contrary, the  probabilities of success of the IRF of item 60 are quite low as far as 1.5, which means that it is a difficult item, but  the probabilities of success for the  best students are high. In fact, the probabilities of success for item 60 for $\theta$ higher than 2 are higher than those of  item 18. Item 28 has high score for PC 1 and low score for PC2, while it has a score of  nearly zero for PC 3 and PC 4. However, item 60 has low scores for PC 1 and PC 4, a high score for PC 2 and nearly zero for PC 3. In other words, it would have been very difficult to guess the extreme representatives of the sample returned by ADA from an analysis of the scores in Figure \ref{pcascores}.

\begin{figure}[ht]
\centering
\includegraphics[width=0.7\textwidth]{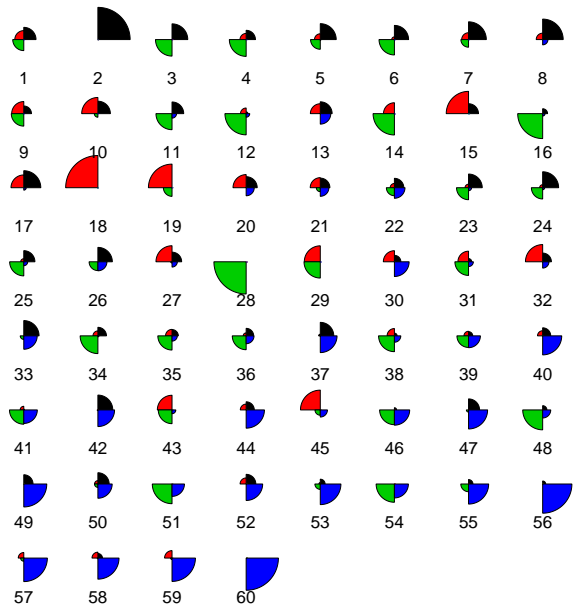}
\caption{\label{alfastarsfig} Star plots of the alphas of each archetypoid for IRFs. The item number appears below each plot. The archetypoids are 2, 18, 28 and 60.}
\end{figure}

The alpha values (from 0 to 1) tell us about the contribution of each archetypoid to each item. Remember that they add up to 1. 
Figure \ref{alfastarsfig} shows star plots of the alpha values for each archetypoid, thus providing a complete human-readable view of the data set. 
 The 4 alpha values in this case are represented starting on the right and going
counter-clockwise around the circle. The size of each alpha is shown by the radius of
the segment representing it. The items that are similar to the archetypoids can be
clearly seen (for example, 7 and 8 are somehow similar to 2; 15 and 19 are somehow similar to 18; 14 and 16 are somehow similar to 28; and 56, 57 and 59 are similar to 60), as can the items that are a mixture of several archetypoids (for example, item 1 is a mixture of mainly item 2, together with items 28 and 18, to a lesser extent). Item 1 was selected by \citet[Ch. 9]{Ramsay02} and \cite{10.2307/3648139} as an example of a low difficulty item, although it seems that item 2 would be a better representative of this kind of item. Item 9 was  selected by \citet[Ch. 9]{Ramsay02} and \cite{10.2307/3648139} as an example of a medium difficulty item, and it is mainly a mixture of items 18 and 28. Finally, item 59 was  selected by \citet[Ch. 9]{Ramsay02} and \cite{10.2307/3648139} as an example of a hard item. Item 59 was mainly explained by item 60.

\textcolor{black}{Results of applying FADA with kernel and parametric IRF estimates are discussed in the Supplementary Material.}



\section{Conclusion}
\label{conclusiones} 

\vspace{-0.2cm}
\textcolor{black}{For the first time, }we have proposed to find archetypal patterns in binary data using ADA for a better understanding of a data set. A simulation study and results provided in two applications have highlighted the benefits of ADA for binary questionnaires as an alternative that can be used instead of (or in addition to) other established methodologies. 

Although, much of statistics is based on the idea that averaging over
many elements of a data set is a good thing to do, in this paper we adopt a different perspective. We have selected a small 
 number of representative observations, archetypal observations, and the
 data composition is explained through mixtures of those extreme observations.  We have shown that this can be highly informative and  is a useful tool for making a data set more
 ``human-readable'', even to non-experts.
 
 \textcolor{black}{In the first application, we have shown how ADA returns the most complementary profiles, which can be more useful in order to establish groups of students with similar mastery of skills. Furthermore, ADA returns composition information of each observation through alpha values, which is a richer information than the simple assignation to groups returned by CLA. In the second application, FADA has discovered the extreme patterns in the data, which cannot be recovered by FPCA. Furthermore, we have explained each item of the ACT math exam as a percentage of the archetypal items, which is easily understandable even for non-experts.}

As regards future work, throughout the paper all variables share the same weight, but for certain situations some variables could have more weight in RSS. 
Another direction of future work would be to consider ADA for nominal observations, for example, by converting those variables into dummy variables, i.e. with binary codes. \textcolor{black}{ Furthermore, this work is limited to binary data, but questionnaires can also have Likert-type scale responses. Therefore, archetypal techniques for ordinal data would be very valuable.} Another not so immediate extension,  would be to consider the case of mixed data, with real valued and categorical data, together with missing data. \textcolor{black}{Finally, from the computational point of view, in case of working with a very big data set, the ADA algorithm described in Section \ref{ADA al} could be slow. In that case, a recent alternative implemented in the R package {\bf adamethods} \citep{adamethods} for computing ADA with large data sets could be used.}

\section*{Acknowledgments} 
    This work is supported by the following grants: DPI2017-87333-R from the Spanish Ministry of
Science, Innovation and Universities (AEI/FEDER, EU) and UJI-B2017-13 from
Universitat Jaume I.

\bibliographystyle{chicago}
\bibliography{SORTR1}

\end{document}